\documentclass[twocolumn,reprint,amsmath,amssymb, aps,prl]{revtex4-2}
\usepackage[T1]{fontenc}
\usepackage[latin9]{inputenc}
\makeatletter
\usepackage{float}
\usepackage{physics}
\usepackage{amsfonts}
\usepackage{latexsym}
\usepackage{epsfig}
\usepackage{graphicx}
\usepackage{tikz}
\definecolor{greatblue}{RGB}{40,120,181}
\definecolor{greatred}{RGB}{200,36,35}
\usepackage[colorlinks,linkcolor=greatblue,anchorcolor=blue,citecolor=greatred]{hyperref}

\usepackage{dcolumn}
\usepackage{bm}

\makeatother

\usepackage[english]{babel}

\begin{document}

\title{ Realization of ``ER=EPR''}

\author{Xin Jiang}  
\email{domoki@stu.scu.edu.cn}
\affiliation{College of Physics, Sichuan University, Chengdu, 610065, China}

\author{Peng Wang}
\email{pengw@scu.edu.cn}
\affiliation{College of Physics, Sichuan University, Chengdu, 610065, China}

\author{Houwen Wu}
\email{iverwu@scu.edu.cn}   
\affiliation{College of Physics, Sichuan University, Chengdu, 610065, China}

\author{Haitang Yang}
\email{hyanga@scu.edu.cn}
\affiliation{College of Physics, Sichuan University, Chengdu, 610065, China}

\begin{abstract}
We provide a concrete and computable realization of the $ER=EPR$ conjecture, by deriving the Einstein-Rosen bridge from the quantum entanglement in the thermofield double CFT. The Bekenstein-Hawking entropy of the wormhole is explicitly identified as an entanglement entropy between subsystems of the thermofield double state. Furthermore, our results provide a quantitative verification of Van Raamsdonk's conjecture about spacetime emergence. 
\end{abstract}

\maketitle
\onecolumngrid

\section{Introduction}

Motivated by the AdS/CFT correspondence \cite{Maldacena:1997re,Gubser:1998bc,Witten:1998qj},
it has been frequently suggested that  spacetime geometry can emerge
from  quantum entanglement \cite{Ryu:2006bv,VanRaamsdonk:2010pw,Maldacena:2013xja}.
A particularly intriguing conjecture is that an Einstein-Rosen
(ER) bridge, or wormhole, connecting two black holes, arises from
entanglement between the microstates of these black holes. This idea
is encapsulated in the slogan \textquotedblleft $ER=EPR$\textquotedblright  \cite{Maldacena:2013xja}.
It was suggested that an AdS wormhole is dual to   two entangled CFTs in a thermofield
double state $\vert\text{TFD}\rangle$ \cite{Maldacena:2001kr,VanRaamsdonk:2010pw}.
This raises a fundamental question:\emph{ How the ER bridge is explicitly realized as
the quantum entanglement encoded in $\vert\text{TFD}\rangle$? }

The most important quantity that describes the degree of quantum entanglement
is the entanglement entropy. For  an entangled pure
state $\rho$, the entanglement entropy of a subsystem $A$ is defined
as the von Neumann entropy 
\begin{equation}
S_{\text{vN}}(A:A^c)=-\mathrm{Tr}\rho_{A}\log\rho_{A},
\label{eq:S}
\end{equation}
where the reduced density matrix $\rho_{A}$ is obtained by tracing
out the complement region $A^{c}$, $\rho_{A}=\mathrm{Tr}_{A^{c}}\rho$.
In QFT, when $A$ and its complement $A^c$ are contiguous, more specifically 
$\partial A \cap \partial A^c \not= 0$ \footnote{We appreciate the anonymous referee for pointing out this.},
the  entanglement entropy $S_\mathrm{adj}(A:A^c)$ is UV divergent due to the 
very intense entanglement between neighboring fields.
The subscript ``adj'' stands for adjacent configuration. 
The asymptotic property of $S_\mathrm{adj}(A:A^c)$ 
makes extracting  spacetime geometry very hard \cite{Wang:2017ele,Wang:2018vbw,Wang:2018jva}, 
if not completely impossible.

\begin{figure}[h]
\includegraphics[width=0.5\textwidth]{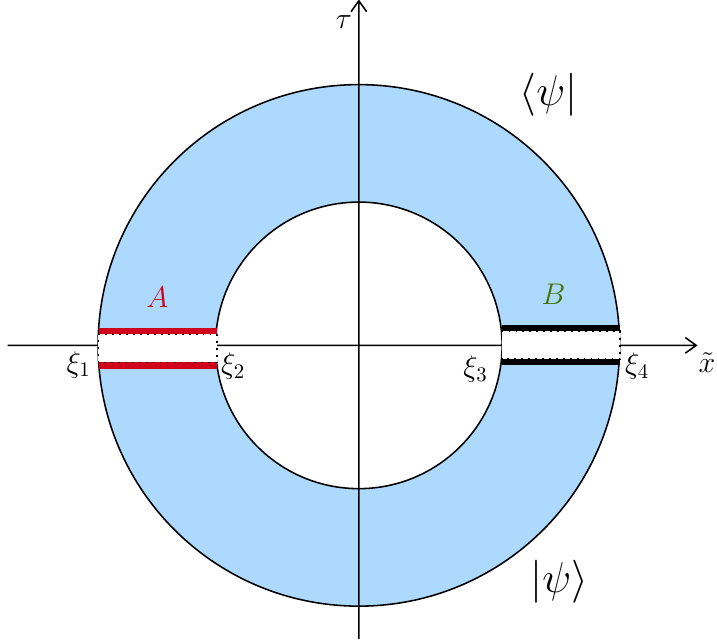}
\caption{The Euclidean pure state density matrix  $\rho =|\psi\rangle \langle\psi|$
for the annular  CFT$_2$.  
Two distinct intervals $A$ and $B$ have finite  entanglement entropy $S_\mathrm{disj}(A:B)$.
\label{fig:density}}
\end{figure}

On the other hand, for UV-complete theories such as CFTs, UV divergences 
should arise as limits of well-defined regular quantities, rather than being 
intrinsic ingredients of the theory itself. 
To this end, let us consider the annulus CFT$_2$  in Figure \ref{fig:density} where
the entangled regions $A$ and $B$ are disjoint, i.e. $\partial A \cap \partial A^c = 0$.
It is crucial that the Euclidean temporal direction is upward. This is clearly a pure state with 
the total density matrix $\rho =|\psi\rangle \langle\psi|$. 
After tracing out $B$, we get the reduced density matrix $\rho_{A}=\mathrm{Tr}_{B}\rho$. 
The entanglement entropy $S_\mathrm{disj}(A:B)$  is then as usual defined by eq. (\ref{eq:S}).
We use the subscript ``disj'' to denote such disjoint configurations.

More precisely, for a planar configuration, \(S_{\rm disj}(A:B)\) is defined through the
subtraction/annular construction.  For a pair of disjoint regions \(A,B\),
one removes the complementary regions \(C,D\) in the Euclidean path
integral.  The resulting annular geometry prepares a pure state
\(|\psi_{AB}\rangle\) on \(\Sigma_{AB}=A\cup B\), where \(A\) and \(B\) are
disjoint but complementary.  We then define
\[
\rho_A^{(AB)}
=
{\rm Tr}_B |\psi_{AB}\rangle\langle\psi_{AB}|,\qquad
S_{\rm disj}(A:B)
=
-{\rm Tr}\rho_A^{(AB)}\log\rho_A^{(AB)} .
\]
When the pair \(A,B\) is embedded in a larger system, such as the TFD
configuration below, \(B\) is therefore not meant to be the complement of
\(A\) in the full Cauchy slice.  The equivalence between this annular
definition and the mixed-state/subtraction interpretation was explained
in detail in the appendix of Ref.~\cite{Jiang:2024hjz}.

In recent works \cite{Jiang:2024ijx,Jiang:2024hjz,Jiang:2025tqu,Jiang:2025dir,Jiang:2025jnk}, 
a systematic method is developed to calculate such disjoint entanglement entropy  $S_{\mathrm{disj}}(A:B)$ 
in Figure \ref{fig:density}. For generic time dependent configurations,
$\xi_i = \tilde x_i + i\tilde t_i$ and $\tilde t_i = -i \tau_i$ for Euclidean time,  we have \cite{Jiang:2025tqu},
\begin{equation}
S_{\text{disj}}(A:B)  =  \frac{c}{12}\log\left(\frac{1+\sqrt{\eta}}{1-\sqrt{\eta}}\right)+\frac{c}{12}\log\left(\frac{1+\sqrt{\bar{\eta}}}{1-\sqrt{\bar{\eta}}}\right),
\label{eq:Asymmetric-EE}
\end{equation}
with the cross ratio
\begin{equation}
\eta=\frac{\xi_{21}\xi_{43}}{\xi_{31}\xi_{42}}, \quad \xi_{ij}\equiv \xi_i -\xi_j.
\label{eq:cross-ratio}
\end{equation}     
As anticipated,  $S_{\mathrm{disj}}(A:B)$ is indeed \emph{UV-complete},
and has adjacent  entanglement entropies $S_{\mathrm{adj}}(A:B)$ as the adjacent limits. 
Now for the symmetric annulus in Figure \ref{fig:density}, 
we use the following linear reparameterization,
\begin{eqnarray}
X &=& \frac{\xi_4 +\xi_1}{2},\quad {X'} = \frac{\xi_3 +\xi_2}{2},\nonumber\\
Z &=& \frac{|\xi_4 -\xi_1|}{2},\quad {Z'} = \frac{|\xi_3 -\xi_2|}{2}.
\label{eq:Sys-Re}
\end{eqnarray}
We  see that $Z$ and $Z'$ are  the inner radius and outer radius respectively,  representing energy scales. 
For generic configurations such as  asymmetrical annuli, 
the reparameterization is obtained by the corresponding conformal maps, which yields,
\begin{eqnarray}
X & = & \xi_{1}+\frac{1}{1+\alpha}\xi_{41},\quad {X^{\prime}}  =  \xi_{2}+\frac{1}{1+\beta}\xi_{32},\\
Z & = & \frac{\sqrt{\alpha}}{1+\alpha}\vert\xi_{14}\vert,\quad\quad Z^{\prime}  =  \frac{\sqrt{\beta}}{\beta+1}\vert\xi_{23}\vert,
\label{eq:Asys-Re}
\end{eqnarray}
with $\alpha=\frac{\vert\xi_{34}\vert\vert\xi_{24}\vert}{\vert\xi_{13}\vert\vert\xi_{12}\vert}$,
$\beta=\frac{\vert\xi_{13}\vert\vert\xi_{34}\vert}{\vert\xi_{12}\vert\vert\xi_{24}\vert}$ and the cross ratio 
\begin{equation}
\eta = \frac{\vert\xi_{12}\vert\vert\xi_{34}\vert}{\vert\xi_{13}\vert\vert\xi_{24}\vert}=
\frac{(Z-Z')^{2}+\vert X-X'\vert^{2}}{(Z+Z')^{2}+\vert X-X'\vert^{2}}.\label{eq:cross-ratio-x}
\end{equation}
We now introduce a  critical grouping by treating the energy scale $Z$, $Z'$ as an extra dimension
\begin{equation}
Y^\mu \equiv (X,Z), \quad {Y'}^\mu\equiv (X',Z'),
\end{equation}
which can be interpreted as two points in a three dimensional spacetime.
After rewriting the entropy  eq. (\ref{eq:Asymmetric-EE}) in terms of $Y^\mu$,
in \cite{Jiang:2024hjz}, 
we showed how three dimensional Einstein's equation emerges from CFT$_2$, and
demonstrated that the dual spacetime metric can be extracted  from  disjoint  entanglement entropies straightforwardly
as follows, 
\begin{equation}
\chi:=\frac{1}{2}S_{\mathrm{disj}}^{2}(A:B),\quad g_{\mu\nu} = -\lim_{Y\to Y'} \partial_{Y^\mu} \partial_{{Y'}^\nu} \chi.
\label{eq:entropic-func}
\end{equation}

The purpose of this paper is to first compute the disjoint entanglement 
entropy $S_{\mathrm{disj}}(A:B)$ of a particular configuration in
$\vert\text{TFD}\rangle$. Then
by employing the entropic function $\chi = \frac{1}{2}S_{\mathrm{disj}}^{2}(A:B)$, 
we are going to derive 
the ER bridge  from the disjoint entanglement entropy $S_{\mathrm{disj}}^{2}(A:B)$.
We also present
a verification of Van Raamsdonk's conjecture \cite{VanRaamsdonk:2010pw}.

\section{Disjoint Entanglement in the $\vert\text{TFD}\rangle$ }

We work with CFT$_2$ living on a spatially infinite line.
The ``thermofield double'' state $\vert\text{TFD}\rangle$ is  
defined by an entangled pure state of two copies of  thermal CFT$_2$:
\begin{equation}
\vert\text{TFD}\rangle=\sum_{n}e^{-\frac{\beta}{2}E_{n}}\vert n_{L}\rangle\otimes\vert n_{R}\rangle.
\end{equation}
Here, $\beta^{-1}$ is the temperature, and $\vert n_{L,R}\rangle$
are the $n$-th energy eigenstates of individual systems. Note that
each copy of CFTs is in the mixed thermal state 
\begin{equation}
e^{-\beta H}=\mathrm{Tr}_{L}\vert\text{TFD}\rangle\langle\text{TFD}\vert=\mathrm{Tr}_{R}\vert\text{TFD}\rangle\langle\text{TFD}\vert.
\end{equation}

\begin{figure}[h]
\centering{}\includegraphics[width=0.5\textwidth]{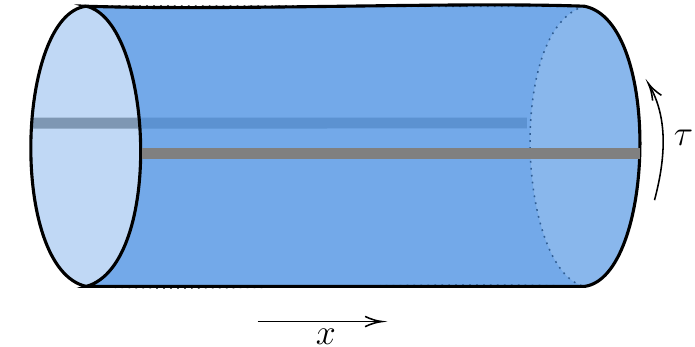}
\caption{The density matrix $\rho=\vert\text{TFD}\rangle\langle\text{TFD}\vert$,
the blue shaded region represents the Euclidean path integral.
The two gray   infinite slits are states
living in left and right Hilbert space respectively.\label{fig:cylinder-1}
We have set   $x\in(-\infty,+\infty)$ and $\tau\in[0,\beta)$.}
\end{figure}

By using the Euclidean path integral, every thermal state $e^{-\beta H}$
can be visually represented as a cylinder with an infinite slit. Therefore,
$\rho=\vert\text{TFD}\rangle\langle\text{TFD}\vert$ is simply a cylinder
with two infinite slits, as shown in Fig. \ref{fig:cylinder-1}.

\begin{figure}[h]
\centering{}\includegraphics[width=0.5\textwidth]{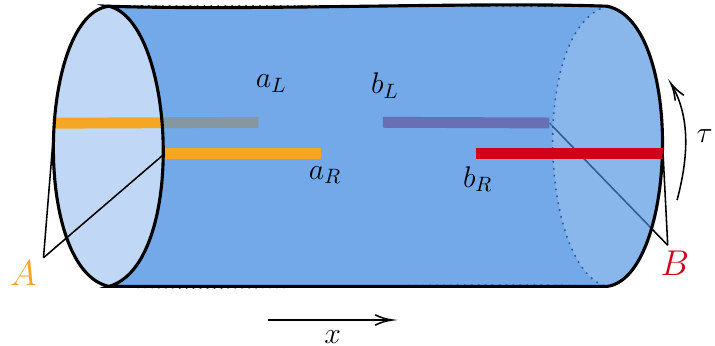}
\caption{At $t=0$ slice, we calculate the disjoint entanglement entropy between the
disconnected segments $A$ and $B$.
\label{fig:EE_in_AB}}
\end{figure}

To extract the dual metric by using eq. (\ref{eq:entropic-func}), 
we need an exact  entanglement entropy between  disjoint regions. 
Particularly, in order to derive ER bridge, we  choose 
sub-region $A$ as  $(-\infty,a_{L})\cup(-\infty,a_{R}),$
and $B$ as $(b_{L},+\infty)\cup(b_{R},+\infty)$ at $t=0$ slice, depicted in Fig. \ref{fig:EE_in_AB}. 
Although \(A\cup B\) is not a complete Cauchy slice of the full TFD
system, the finite quantity \(S_{\rm disj}(A:B)\) is well-defined by the
subtraction/annular prescription recalled above.

Following  \cite{Jiang:2024ijx,Jiang:2024hjz,Jiang:2025tqu,Jiang:2025dir}, 
by applying the conformal transformation
$z=e^{2\pi w/\beta}$ which maps the cylinder to the complex plane,
using the entanglement entropy eq. (\ref{eq:Asymmetric-EE}) for $\tau=0$  slice,  
it is straightforward to obtain
the entanglement entropy  between $A$ and $B$
\begin{equation}
S_{\text{disj}}(A:B)=\frac{c}{6}\log\left[1+2\eta+2\sqrt{\eta(\eta+1)}\right],
\end{equation}
where $c$ is the central charge and the cross-ratio is
\begin{equation}
\eta=\frac{\cosh(\frac{\pi(a_{L}-a_{R})}{\beta})\cosh(\frac{\pi(b_{L}-b_{R})}{\beta})}{\sinh(\frac{\pi(a_{L}-b_{L})}{\beta})\sinh(\frac{\pi(a_{R}-b_{R})}{\beta})}.
\end{equation}
We see  $S_{\text{disj}}(A:B)$ is  exact and UV-finite,
fully quantifying the entanglement between $A$ and $B$. 
Now applying the conformal transformation
$z=e^{2\pi w/\beta}$
again on the reparameterization eq. (\ref{eq:Asys-Re}),
it is straightforward to derive the following  reparameterization for the cylinder
\begin{eqnarray}
\frac{2u}{1+u^{2}}e^{\phi} & = & -\frac{2e^{\frac{2\pi(a_{L}+a_{R}+b_{L})}{\beta}}+e^{\frac{2\pi(a_{L}+a_{R}+b_{R})}{\beta}}+e^{\frac{2\pi(2a_{L}+b_{L})}{\beta}}+e^{\frac{2\pi(a_{L}+2b_{L})}{\beta}}+2e^{\frac{2\pi(a_{L}+b_{L}+b_{R})}{\beta}}+e^{\frac{2\pi(a_{R}+b_{L}+b_{R})}{\beta}}}{e^{\frac{2\pi(a_{L}+a_{R})}{\beta}}+e^{\frac{4\pi a_{L}}{\beta}}+e^{\frac{2\pi(a_{L}+b_{R})}{\beta}}+e^{\frac{2\pi(a_{R}+b_{L})}{\beta}}+2e^{\frac{2\pi(a_{R}+b_{R})}{\beta}}+e^{\frac{4\pi b_{L}}{\beta}}+e^{\frac{2\pi(b_{L}+b_{R})}{\beta}}},\nonumber \\
\frac{1-u^{2}}{1+u^{2}}e^{\phi} & = & \frac{4\left(e^{\frac{2\pi b_{L}}{\beta}}-e^{\frac{2\pi a_{L}}{\beta}}\right)e^{\frac{\pi(a_{L}+a_{R}+b_{L}+b_{R})}{\beta}}\Upsilon}{e^{\frac{2\pi(a_{L}+a_{R})}{\beta}}+e^{\frac{4\pi a_{L}}{\beta}}+e^{\frac{2\pi(a_{L}+b_{R})}{\beta}}+e^{\frac{2\pi(a_{R}+b_{L})}{\beta}}+2e^{\frac{2\pi(a_{R}+b_{R})}{\beta}}+e^{\frac{4\pi b_{L}}{\beta}}+e^{\frac{2\pi(b_{L}+b_{R})}{\beta}}},\nonumber \\
\frac{2u^{\prime}}{1+u^{\prime2}}e^{\phi^{\prime}} & = & \frac{2e^{\frac{2\pi(a_{L}+a_{R}+b_{L})}{\beta}}+e^{\frac{2\pi(a_{L}+a_{R}+b_{R})}{\beta}}+e^{\frac{2\pi(2a_{R}+b_{R})}{\beta}}+e^{\frac{2\pi(a_{R}+2b_{R})}{\beta}}+2e^{\frac{2\pi(a_{L}+b_{L}+b_{R})}{\beta}}+e^{\frac{2\pi(a_{R}+b_{L}+b_{R})}{\beta}}}{e^{\frac{2\pi(a_{L}+a_{R})}{\beta}}+e^{\frac{4\pi a_{R}}{\beta}}+e^{\frac{2\pi(a_{L}+b_{R})}{\beta}}+e^{\frac{2\pi(a_{R}+b_{L})}{\beta}}+2e^{\frac{2\pi(a_{R}+b_{R})}{\beta}}+e^{\frac{4\pi b_{R}}{\beta}}+e^{\frac{2\pi(b_{L}+b_{R})}{\beta}}},\nonumber \\
\frac{1-u^{\prime2}}{1+u^{\prime2}}e^{\phi^{\prime}} & = & \frac{4\left(e^{\frac{2\pi b_{R}}{\beta}}-e^{\frac{2\pi a_{R}}{\beta}}\right)e^{\frac{\pi(a_{L}+a_{R}+b_{L}+b_{R})}{\beta}}\Upsilon}{e^{\frac{2\pi(a_{L}+a_{R})}{\beta}}+e^{\frac{4\pi a_{R}}{\beta}}+e^{\frac{2\pi(a_{L}+b_{R})}{\beta}}+e^{\frac{2\pi(a_{R}+b_{L})}{\beta}}+2e^{\frac{2\pi(a_{R}+b_{R})}{\beta}}+e^{\frac{4\pi b_{R}}{\beta}}+e^{\frac{2\pi(b_{L}+b_{R})}{\beta}}},
\label{cylinder-repara}
\end{eqnarray}
where
\begin{equation}
\Upsilon=\sqrt{\cosh\left(\frac{\pi(a_{L}-a_{R})}{\beta}\right)\cosh\left(\frac{\pi(a_{L}-b_{R})}{\beta}\right)\cosh\left(\frac{\pi(a_{R}-b_{L})}{\beta}\right)\cosh\left(\frac{\pi(b_{L}-b_{R})}{\beta}\right)}.
\end{equation}
$S_{\text{disj}}(A:B)$  can then be put into
\begin{equation}
S_{\text{disj}}(A:B)=\frac{c}{6}\text{arccosh}\left(\frac{\left(u^{2}+1\right)\left(u^{\prime2}+1\right)\cosh\left(\phi^{\prime}-\phi\right)-4uu^{\prime}}{\left(u^{2}-1\right)\left(u^{\prime2}-1\right)}\right).
\label{eq:AB-EE}
\end{equation}

\section{ER $=$ EPR}

Now, we apply eq. (\ref{eq:entropic-func}) with $x^\mu=(u,\phi)$ and ${x'}^\mu=(u',\phi')$ to get 
the    entanglement entropy induced spacetime metric

\begin{eqnarray}
ds^2 &&=\Big[ -\lim_{x'\to x}\partial_{x'^{i}}\partial_{x^{j}}\Big(\frac{1}{2}S^{2}_{\text{disj}}(A:B)\Big)\Big] dx^i dx^j\nonumber \\
&&= \frac{4}{(1-u^{2})^{2}}du^2 +  \left(\frac{1+u^{2}}{1-u^{2}}\right)^{2} d\phi^2,
\label{eq:ER-Metric}
\end{eqnarray}
where for simplicity, we set $c=6$. The coordinate ranges are
	\begin{equation}
		-1\le u\le 1,\qquad -\infty<\phi<\infty .
	\end{equation}
	These follow directly from the reparameterization. Indeed, 
	for simplicity, setting
	\begin{equation}
		a_L=a_R=-b_L=-b_R=-b\le 0,
		\label{eq:ab-1}
	\end{equation}
	gives
	\begin{equation}
		u=-e^{-2\pi b/\beta},\qquad u'=e^{-2\pi b/\beta},
	\end{equation}
	from which \(|u|\le 1\) is immediate. Other values of $a_L$, $a_R$, $b_L$ and $b_R$ can 
	be achieved by conformal transformations. 
	Similarly, setting
	\begin{equation}
		a_L=a_R=b_L=b_R=x,
		\label{eq:ab-1}
	\end{equation}
	yields
	\begin{equation}
		\phi=\frac{2\pi x}{\beta},
	\end{equation}
	so \(\phi\) ranges over the full real line as \(x\in\mathbb{R}\), 
	obtained by unrolling a periodic angular coordinate.
Remarkably, this metric reveals
a geometry in which two spatial subregions ($u<0$ and $u>0$) are
connected by a wormhole throat located at $u=0$, with the unit radius. 

\begin{figure}[h]
\centering{}\includegraphics[width=0.4\textwidth]{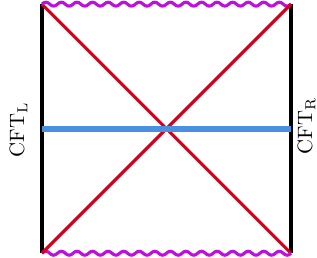}
\caption{Penrose diagram of the eternal AdS black hole. Two vertical
lines represent two CFTs living on the left and right boundaries respectively.
The blue line represents the time slice $T=0$.\label{fig:Penrose-diagram-of}}
\end{figure}
To see it more explicitly, we refer to the eternal black hole in Kruskal coordinates \cite{Maldacena:2001kr}:
\begin{equation}
ds^{2}=-\frac{4}{(1+uv)^{2}}dudv+\left(\frac{1-uv}{1+uv}\right)^{2}d\phi^{2},
\label{eq:ER-Kruskal}
\end{equation}
which includes a non-traversable wormhole connecting the left and right regions, as depicted in Fig.
\ref{fig:Penrose-diagram-of}. 
Since $u=T+R$, $v=T-R$, at $T=0$, one has $u=-v$, and the eternal black hole becomes precisely the entanglement  entropy induced  metric (\ref{eq:ER-Metric}).

\begin{figure}[h]
\centering{}\includegraphics[width=0.4\textwidth]{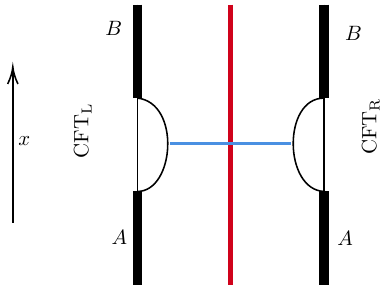}
\caption{The red line represents the horizon. The blue line indicates
the RT surface of the entanglement entropy between
$A$ and $B$. \label{fig:RT-surface}}
\end{figure}

Holographically, from the Ryu-Takayanagi formula,
the gravity dual of the disjoint entanglement entropy $S_{\text{disj}}(A:B)$
is the blue geodesic living in the ER bridge at  $T=0$,
as schematically depicted in Fig. \ref{fig:RT-surface}. 
It is usually termed as entanglement wedge cross-section (EWCS) 
with finite length which exactly equals the 
disjoint entanglement entropy $S_{\text{disj}}(A:B)$.

We thus  have successfully derived the ER bridge from the
entanglement entropy of  $\vert\text{TFD}\rangle$.

\section{Bekenstein-Hawking Entropy of wormhole as Entanglement Entropy}

\begin{figure}[h]
\centering{}\includegraphics[width=0.9\textwidth]{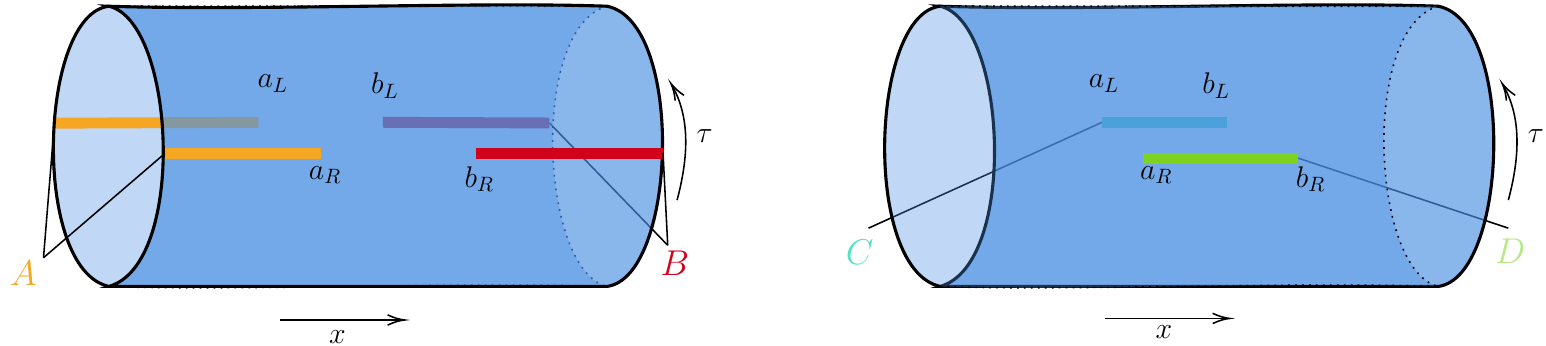}
\caption{Segments $C$, $D$ and  $A$,  $B$  represent two different disjoint configurations 
in the entangled TFD state. \label{fig:EE_in_TFD}}
\end{figure}

\begin{figure}[h]
\includegraphics[width=1\textwidth]{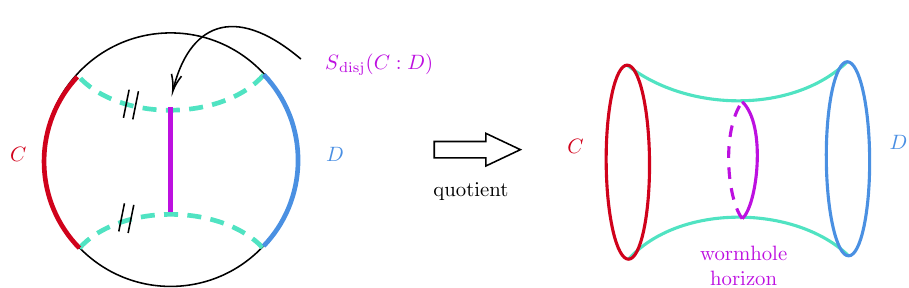}
\centering{}\caption{Left panel:  An entanglement wedge is bounded by two cyan
geodesics. The purple line is the EWCS. Right panel: A wormhole can be prepared
by identifying two cyan geodesics of the entanglement wedge.
\label{fig:quotient}}
\end{figure}

In addition to  $S_{\mathrm{disj}}\left(A:B\right)$ addressed in the preceding section, referring to 
Fig. \ref{fig:EE_in_TFD}, let us consider the entanglement between the disconnected segments $C$ and $D$,
which separate $A$ and $B$ at $t=0$ slice. 
As shown in \cite{Jiang:2024ijx,Jiang:2025dir}, one gets
\begin{equation}
S_{\text{disj}}(C:D)=\frac{c}{6}\log\left[1+2\eta^{-1}+2\sqrt{\eta^{-1}(\eta^{-1}+1)}\right].
\end{equation}
Referring to Figure \ref{fig:quotient},
in the bulk geometry, $S_{\mathrm{disj}}\left(C:D\right)$ admits two
distinct interpretations:
\begin{itemize}
\item \textbf{Entanglement Wedge Interpretation:} 
The map is the standard conformal map from the thermal cylinder to the
complex plane.  Writing \(w=x+i\tau\) with \(\tau\sim\tau+\beta\), the
exponential map
\[
z=e^{2\pi w/\beta}
\]
maps the Euclidean thermal cylinder to the punctured complex plane.  After
including the point \(z=\infty\), the plane is conformally compactified to
a Riemann sphere; equivalently, on a constant time slice the spatial line
is compactified to a circle.  Under this compactification the intervals
\(C\) and \(D\) become two boundary intervals on the circle.  Their
entanglement wedge is the bulk region bounded by the corresponding RT
geodesics, as shown in the left panel of Fig.~7, and the minimal geodesic
connecting these two RT geodesics is the EWCS.
It is widely accepted
that the spacetime subregion associated with the entanglement between
$C$ and $D$ is the entanglement wedge \cite{Czech:2012bh,Wall:2012uf,Headrick:2014cta},
and $S_{\mathrm{disj}}\left(C:D\right)=E_W$, the purple geodesic, referred as EWCS   
\cite{Jiang:2024ijx,Takayanagi:2017knl}.
\item \textbf{Wormhole Interpretation:} In the context of wormhole geometry,
it is well-established that a two-sided wormhole can be constructed
by identifying two cyan geodesics \cite{Brill:1998pr}. The minimal
surface connecting these cyan geodesics corresponds to the horizon
area of the wormhole \cite{Bao:2018fso}.
\end{itemize}
These two interpretations indicate that the EWCS is equal to the horizon area of the
wormhole. More importantly, the entanglement entropy $S_{\mathrm{disj}}\left(C:D\right)$
equals the Bekenstein-Hawking entropy of the wormhole.

\noindent However, previous studies lack a proper configuration to 
verify this equivalence. Let us now use our results to address
this agreement. By setting $a_{L}=a_{R}=a$ and $b_{L}=b_{R}=b$,
the entanglement entropy $S_{\text{disj}}(C:D)$ simplifies to a thermal
entropy,
\begin{eqnarray}
S_{\text{disj}}(C:D) & = & \frac{\pi c}{3\beta}\vert a-b\vert,\nonumber\\
 & = & \frac{2\pi r_{+}}{4G^{(3)}}.
\label{eq:WormHorizon}
\end{eqnarray}
The second line follows from  the parameter relations in AdS/CFT,  $c=3\ell/2G^{(3)}$  and $\frac{|a-b|}{\beta} = \frac{r_+}{\ell}$,
with the Hawking temperature $\beta = \frac{2\pi\ell^2}{r_+}$. In the gravity, the wormhole horizon is simply a kind of EWCS. What our construction adds is  an explicit CFT realization of  this relation: the entanglement entropy $S_{\mathrm{disj}}(C:D)$ reproduces the EWCS and, in the symmetric limit, gives precisely the Bekenstein--Hawking entropy.
This \textit{symmetric} entanglement entropy $S_{\mathrm{disj}}(C:D)$ is in perfect agreement with the Bekenstein-Hawking entropy of the wormhole:
\begin{equation}
S_{\text{disj}}(C:D) =\frac{\mathsf{A}}{4G_{N}}=S_{\mathrm{BH}},
\end{equation}
\noindent where $\mathsf{A}=2\pi r_{+}$ denotes the horizon area
at the throat of the wormhole. 

\section{Verification of Van Raamsdonk's conjecture}
\begin{figure}[h]
\includegraphics[width=1\textwidth]{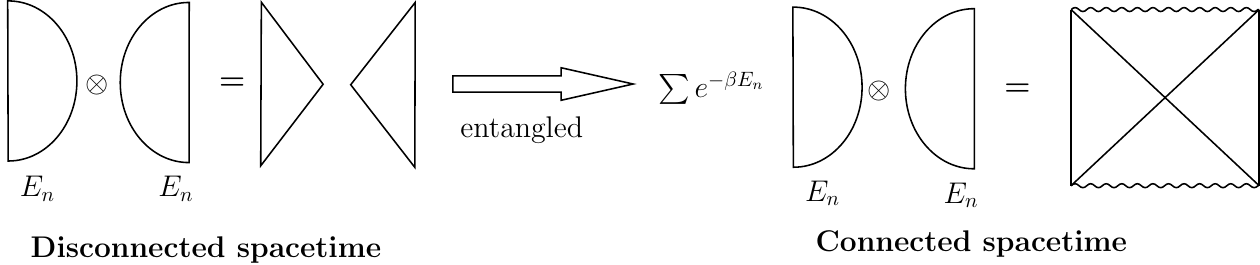}
\centering{}\caption{The connected spacetime can be created by quantum superposition of   disconnected spacetimes.\label{fig:Raamsdonk picture}}
\end{figure}

Our calculations provide a verification to Van  Raamsdonk's conjecture \cite{VanRaamsdonk:2010pw}
that classically connected spacetime emerges from quantum entanglement.
To understand this, let us briefly review Van Raamsdonk's thought experiment.
Referring to Fig. \ref{fig:Raamsdonk picture},
when two subsystems are not entangled, their combined state is a product state: $\left|\Psi\right\rangle =\left|\Psi_{1}\right\rangle \otimes\left|\Psi_{2}\right\rangle $.
In the framework of AdS/CFT, this state corresponds
to two disconnected spacetimes, with $\Psi_{i}$ representing the
$i$-th dual spacetime. As the two states become entangled, the product
state transitions to a TFD state:

\begin{equation}
\left|\mathrm{TFD}\right\rangle =\underset{n}{\sum}e^{-\frac{\beta}{2}E_{n}}\left|n_{L}\right\rangle \otimes\left|n_{R}\right\rangle.
\end{equation}

\noindent So the $\vert\text{TFD}\rangle$ represents the quantum superposition
of disconnected spacetimes, giving rise to a connected spacetime,
such as a wormhole.

\begin{figure}[h]
\begin{centering}
\includegraphics[width=1\textwidth]{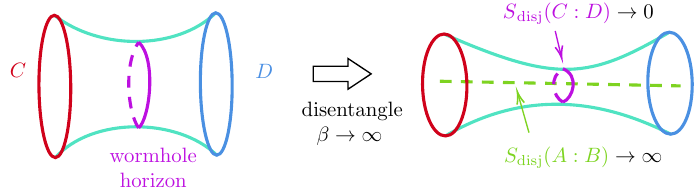}
\par\end{centering}
\centering{}\caption{Disentangle the degrees of freedom in $C$ and $D$ by decreasing
the temperature. The proper length between the corresponding spacetime
regions increases to infinity, while the horizon area decreases to
zero. \label{fig:Raamsdonk verification}}
\end{figure}

In the limit  $\beta\rightarrow\infty$, the entangled
$\vert\text{TFD}\rangle$  reduces back to the unentangled product state ($E_0 =0$), and the wormhole
disappears. In this case, the wormhole horizon area approaches zero,
and its proper length diverges, as argued by Van  Raamsdonk and illustrated
in Fig. \ref{fig:Raamsdonk verification}.

Our results offer a direct verification of this conjecture. 
The entanglement
entropy $S_{\mathrm{disj}}\left(C:D\right)$ reflecting the wormhole horizon
area from eq. (\ref{eq:WormHorizon}),  vanishes as $\beta\rightarrow\infty$:

\begin{equation}
S_{\mathrm{disj}}\left(C:D\right)=\frac{\pi c}{3\beta}\left|a-b\right|\rightarrow0.
\end{equation}

\noindent Whereas the entanglement entropy $S_{\mathrm{disj}}\left(A:B\right)$
corresponding to the wormhole ``length'',  diverges as $\beta\rightarrow\infty$:

\begin{equation}
S_{\mathrm{disj}}\left(A:B\right)=\frac{c}{6}\log\left[1+2\eta+2\sqrt{\eta\left(\eta+1\right)}\right]\sim\frac{c}{3}\log\beta\rightarrow\infty,
\end{equation}

\noindent 
exactly as Van  Raamsdonk conjectured.
It is of importance to note,
in our calculations, the entangled regions $A$ and $B$ (or $C$
and $D$) remain fixed. This guarantees that the reduction in entanglement
entropy arises only from the disentanglement of the two regions, rather
than from a decrease in the size of the entangled regions.

\subsection{Summary}

We have shown that the ER bridge can be derived directly from quantum
entanglement between two CFTs in a TFD state. This result provides
a concrete and computable realization of the $ER=EPR$ conjecture,
and explicitly links quantum entanglement with the emergent ER bridge.
Our results  indicate that the entropic function $\chi=\frac{1}{2}S_{\text{vN}}^{2}$ may
serve as an important tool for probing geometric structures in more
complex entangled states. Furthermore, we showed that the entanglement
wedge cross-section is equivalent to the horizon area of a wormhole.
We then quantitatively verified Van Raamsdonk's conjecture,
supporting the idea that classically connected spacetime arises from
quantum entanglement.

\subsubsection*{Acknowledgements}

This work is supported in part by NSFC (Grant No. 12105191, 12275183
and 12275184).

\appendix

\bibliographystyle{unsrturl}
\bibliography{ref202601}

\end{document}